\documentclass[compsoc, conference, a4paper, 10pt, times]{IEEEtran}

\usepackage{cite}
\usepackage{amsmath,amssymb,amsfonts}
\usepackage{algorithmic}
\usepackage{graphicx}
\usepackage{textcomp}
\usepackage{xcolor}
\usepackage{booktabs}
\usepackage[hidelinks]{hyperref}

\usepackage{caption}
\usepackage{algorithm}
\usepackage{subcaption}
\usepackage{multirow}
\usepackage{tabularx} 
\usepackage{comment}
\usepackage{xspace}
\usepackage{makecell}
\usepackage{mathtools}
\usepackage{enumitem}
\usepackage{tikz}
\usepackage{url}
\usepackage{newfloat}
\usepackage{listings}
\usepackage{bibentry}
\usepackage{dsfont}
\usepackage{flushend}

\begin{document}

\title{Dynamic Cluster Analysis to Detect and Track Novelty in Network Telescopes}

\author{\IEEEauthorblockN{Kai Huang}
\IEEEauthorblockA{\textit{Politecnico di Torino} \\
Turin, Italy \\
kai.huang@polito.it}
\and
\IEEEauthorblockN{Luca Gioacchini}
\IEEEauthorblockA{\textit{Politecnico di Torino} \\
Turin, Italy \\
luca.gioacchini@polito.it}
\and
\IEEEauthorblockN{Marco Mellia}
\IEEEauthorblockA{\textit{Politecnico di Torino} \\
Turin, Italy \\
marco.mellia@polito.it}
\and
\IEEEauthorblockN{Luca Vassio}
\IEEEauthorblockA{\textit{Politecnico di Torino} \\
Turin, Italy \\
luca.vassio@polito.it}
}

\newcommand{\mm}[1]{{\color{red}{[Mellia: #1]}}}
\newcommand{\lgg}[1]{{\color{orange}{[LG: #1]}}}
\newcommand{\kh}[1]{{\color{cyan}{[Kai: #1]}}}
\newcommand{\lv}[1]{{\color{blue}{[LV: #1]}}}

\newcommand{\ie}{\mbox{i.e.}\xspace}
\newcommand{\eg}{\mbox{e.g.}\xspace}

\maketitle

\begin{abstract}


In the context of cybersecurity, tracking the activities of coordinated hosts over time is a daunting task because both participants and their behaviours evolve at a fast pace.
We address this scenario by solving a dynamic novelty discovery problem with the aim of both re-identifying patterns seen in the past and highlighting new patterns. We focus on traffic collected by Network Telescopes, a primary and noisy source for cybersecurity analysis. 
We propose a 3-stage pipeline: (i) we learn compact representations (embeddings) of hosts through their traffic in a self-supervised fashion; (ii) via clustering, we distinguish groups of hosts performing similar activities; (iii) we track the cluster temporal evolution to highlight novel patterns.
We apply our methodology to 20 days of telescope traffic during which we observe more than 8 thousand active hosts.
Our results show that we efficiently identify 50-70 well-shaped clusters per day, 60-70\% of which we associate with already analysed cases, while we pinpoint 10-20 previously unseen clusters per day. These correspond to activity changes and new incidents, of which we document some.
In short, our novelty discovery methodology enormously simplifies the manual analysis the security analysts have to conduct to gain insights to interpret novel coordinated activities.

\end{abstract}


\section{Introduction}
\label{s:intro}

In the context of network monitoring and cybersecurity, the ever-evolving landscape presents a perpetual challenge, with the daily emergence of new threats and vulnerabilities~\cite{kumari2023survey,sun2023survey}. Take botnets which have always attracted the attention of network administrators and security experts. Botnets are groups of machines compromised by the same malware acting under the control of a central server (Command-and-Control paradigm~\cite{padhiar2023c2c}). These compromised machines become part of a large system that the controller can remotely instruct to engage in coordinated malicious activities. 

Botnets constantly evolve, adapting their attack patterns and network scanning behaviours to the latest vulnerabilities. For instance, the FritzFrog botnet~\cite{fritzfrog2022} started exploiting the Log4Shell vulnerability~\cite{wang2023exploring} to improve its spreading strategy~\cite{frog4shell2024}. Similarly,
the widely known Mirai botnet~\cite{affinito2023evolution,antonakakis2017understanding} evolved to at least three new variants only in 2023~\cite{miraivariant2023a,
miraivariant2023b}.
This dynamic nature requires a proactive and collaborative cybersecurity approach for effective mitigation and eradication~\cite{quakbot2023}.

\emph{How to represent and detect such coordinated activities} among senders is a crucial aspect of cyber defence.
Solutions based on Artificial Intelligence (AI) have proven instrumental in identifying these coordinated groups of senders,
offering a proactive approach to cybersecurity. Many works represent senders through traditional feature engineering approaches and process the resulting datasets through (sparse) Autoencoders~\cite{lotfollahi2017deep,hochst2017clustering,KallitsisMerit}, traditional Convolutional Neural Networks~\cite{rezaei2018networking,aceto2019mimetic} or (temporal) Graph Neural Networks~\cite{gioacchini2023exploring}. Other works identify analogies between sequences of packets (or flows) and words in text documents. Hence, they represent senders borrowing techniques from the Natural Language Processing (NLP) field~\cite{ring2017ip2vec,cohen_dante_2020,gioacchini2021darkvec,gioacchini2023idarkvec}.
The detection of senders performing similar actions is then faced by addressing a clustering problem through unsupervised techniques in the embedding space~\cite{hochst2017clustering,gioacchini2021darkvec,gioacchini2023idarkvec}.

However, \emph{distinguishing clusters and activity patterns that are already known from those that are novel} is still an open challenge. Consider the clustering results obtained from analysing data referring to two periods of time, e.g., yesterday and today. Dynamic Cluster Analysis (DCA) aims at identifying which of today's clusters are novel and which can be mapped to yesterday's activity.
Among many evolutionary clustering techniques for data stream~\cite{mahdi2021scalable,liu2022skdstream,maia2020evolving},  MONIC~\cite{spiliopoulou2013monic} is one of the most adopted solutions. 
It defines a framework for modelling and tracking cluster transitions over time. In short, it defines a set of possible evolutions to describe cluster dynamics and to match clusters across temporal snapshots. 

Clustering techniques are commonly used to analyse network telescope traffic, with embeddings \cite{gioacchini2021darkvec}, \cite{KallitsisMerit}, leveraging feature vectors \cite{iglesias2017pattern}, \cite{cabana2021threat}, or based on graphs \cite{soro2020sensing}. However, very few works monitor the evolution of activities and detect novelties. The closest to our work is \cite{KallitsisMerit}, in which the authors proposed an optimal mass transport problem to detect temporal changes in clusters, whereas their approach focuses on unveiling large-scale incidences like Mirai onset in 2016 but lacks tracking the evolution of individual clusters and identifying newly emerged activities.

In this paper, we address these challenges by solving a dynamic novelty discovery problem focusing on Network Telescopes, or Darknets. They are subnets that host neither production services nor clients~\cite{fachkha_darknet_2016} and only observe received unsolicited traffic. They represent a privileged source of information for network security and monitoring activities~\cite{fachkha_inferring_2015, jonker_millions_2017}.
Here, we analyse 20 days of telescope traffic, i.e., 20 independent snapshots of data. We automatically detect groups of senders exhibiting coordination in the traffic they send and study the evolution and dynamics over time of such groups.

\begin{figure}[t]
    \centering
    \includegraphics[width=.8\linewidth]{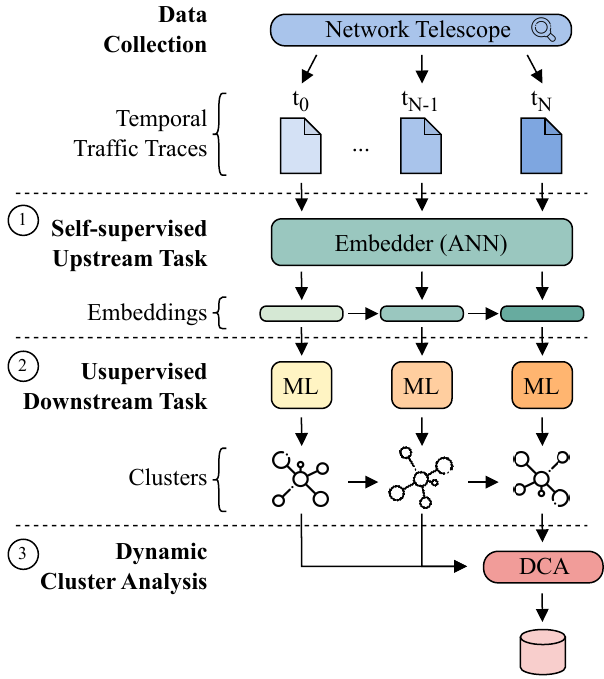}
    \caption{3-stage pipeline for dynamic novelty discovery.}
    \label{fig:pipeline}
\end{figure}

In short, we start the known 2-stage pipeline~\cite{liu2022graph,zied2022multimodal,gioacchini2023idarkvec} in which, given a batch of data, a \textit{self-supervised upstream task} (Stage 1) generates compact representations of input data in a latent space (\ie embeddings) without the need for ground truth or prior knowledge on the data; a clustering algorithm (Stage 2) identifies groups of senders exhibiting similar patterns on top of the generated embeddings. Here, we introduce a novel DCA stage (Stage 3) where we match the current clusters with previously seen ones. This allows us to investigate evolving dynamics, and pinpoint new previously unseen events.

Our analysis shows that the proposed approach greatly simplifies the analysis of the humongous amount of data a network telescope collects. Each day, we identify 50-70 clusters of coordinated senders. We immediately match 60-70\% of them to previously known clusters and identify only 10-20 new clusters, each collecting hundreds of senders that exhibit a clear pattern. We manually analyse some of them, identifying some periodic network scans, the sudden emergence of attacks, morphing patterns in a group of senders, etc.

In summary, the key contributions of this work are as follows: We proposed a DCA approach for detecting and tracking coordinated scanning activities on the internet by presenting a novel framework based on MONIC. We apply our approach to 20 days of real-world data collected from a /24 network telescope. The results show our methodology successfully detects novel activities, limiting human effort. We also manually analyse some of the identified cases and report our findings.

All in all, our proposed pipeline demonstrates how modern machine-learning approaches can support the analysis of traffic for cybersecurity goals. In the future, we plan to move into the real-time implementation of the system and continuous analysis.

\section{Network Traffic Analysis Pipeline}
\label{s:pipeline}

Figure~\ref{fig:pipeline} shows the 3-stage pipeline we detail in the following. Starting from the data collection, we summarise the self-supervised construction of embedding (1) and the unsupervised cluster creation (2) that we borrow from previous works. We then present the details of the new DCA process (3) we introduce in this paper.

\begin{table}[!t]
\centering
\footnotesize
\caption{Traffic from active senders data characterisation. Total refers to distinct senders and ports.}

\begin{tabular}{l|cc|cc}
\toprule
\textbf{} & \multicolumn{2}{c|}{\textbf{20 Days}} & \multicolumn{2}{c}{\textbf{Avg. Daily$\pm$std.}} \\
\textbf{} & \makecell{\centering\textbf{IPs}} & \makecell{\centering\textbf{Ports}}  & {\textbf{IPs}} & {\textbf{Ports}} \\
\midrule
\textbf{TCP} & 48,275 & 65,536 & 7,146$\pm$376 & 32,149$\pm$9,367\\
\textbf{UDP} & 5,825 &  3,897 & 1,413$\pm$105 & 413$\pm$105\\
\textbf{ICMP} & 930 & -- & 151$\pm$42 &  --\\
\textbf{GRE} & 1,396 & -- & 286$\pm$154 &  --\\
\midrule
\textbf{Total} & 52,418 & 65,536& 8,088$\pm$443 & 32,294$\pm$9,326 \\
\bottomrule
\end{tabular}
\label{tab:characterization}
\end{table}

\subsection{Network Telescope Sensor}
\label{ss:stage0}

\begin{figure}[!t]
    \centering
    \includegraphics[width=.8\linewidth]{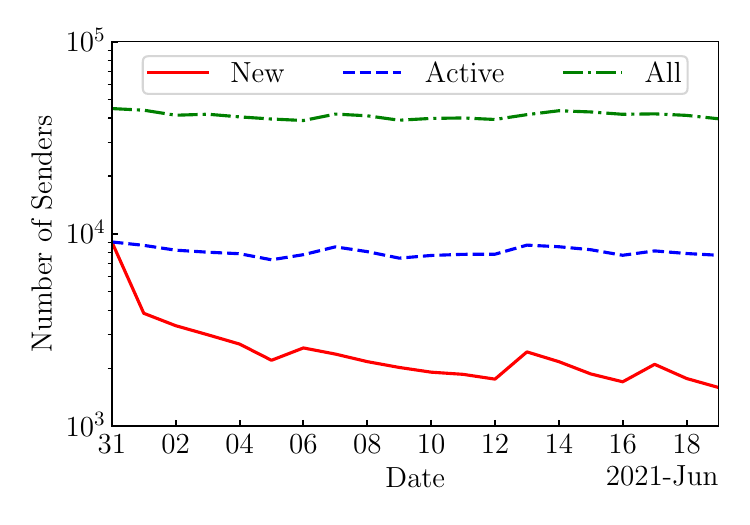}
    \caption{Daily senders on the 20-day dataset.}
    \label{fig:daily_senders}
\end{figure}

\begin{figure}[!t]
    \centering
    \includegraphics[width=.8\linewidth]{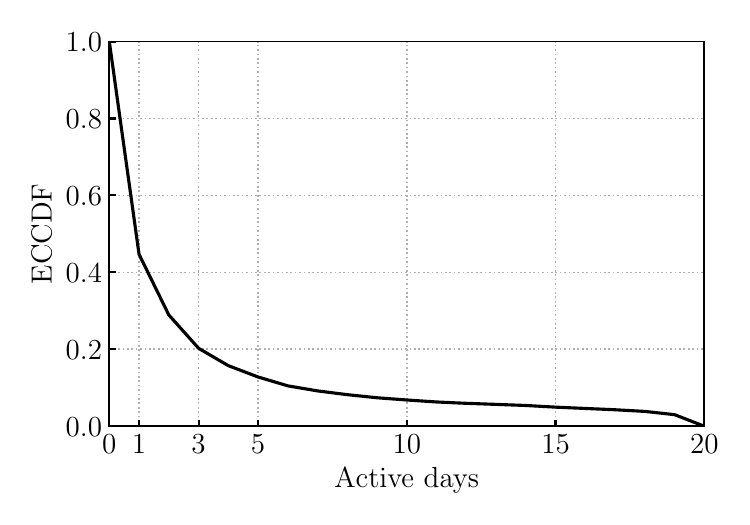}
    \caption{ECCDF of number of active days per sender on the 20-day dataset. Only 20\% of the senders are active for 3 days or more.}
    \label{fig:eccdf_senders}
\end{figure}

In this work, we rely on data collected from the same /24 telescope as in~\cite{gioacchini2023idarkvec} situated in our campus network. We focus on 50 days of telescope traffic, from 2021-05-01 to 2021-06-19. In total, we observed more than {100 million} packets sent by {785 thousand} senders. 
 Here we consider senders sending more than 5 packets daily as ``active''.
After filtering out the inactive senders we are left with about 130,000 active senders in the whole period. We use the first 30 days for bootstrapping the system and the remaining 20 days for testing it. 
Figure~\ref{fig:daily_senders} shows the total number of senders seen each day (green) (about 45,000). About 8,000 are \emph{active} each day (blue), with about 2,000 \emph{new} (red) daily senders appearing for the first time even after 20 days. In fact, most senders are active for very few days as the Empirical Complementary Cumulative Density Function (ECCDF) shows in Figure~\ref{fig:eccdf_senders}.

We detail some further statistics in Table~\ref{tab:characterization}.
In short, we face a very variable scenario, with 52,418 active senders hitting the network telescope in the last 20 days, of which each day $\approx$8,000 are active. They target all TCP and thousands UDP ports, but also generate a significant fraction of ICMP and GRE traffic.

\subsection{Self-Supervised Upstream Task: Sender Embeddings}
\label{ss:stage1}

We generate sender embeddings through the state-of-the-art approach i-DarkVec that we proposed~\cite{gioacchini2023idarkvec}, which relies on Word2Vec~\cite{mikolov2013efficient} and incremental training to automatically create a vector representation for each sender sending traffic to the traffic telescope in a given period. We assume that the reader is familiar with Word2Vec and provide a brief overview of i-DarkVec. 
i-DarkVec receives the sequence of packets as observed by the telescope in one day. Its goal is to represent senders (identified by their IP addresses) so that senders performing similar activities in the same period get projected in the same portion of the space. For this, i-Darkvec extracts the sequences of packets having the same protocol (ICMP/GRE/UDP/TCP) and set of destination ports for TCP/UDP packets.
Analogously to NLP, senders represent ``words'' and their sequence represents ``sentences". We feed the generated sentences as input to Word2Vec to map each sender to a vector in an $E$-dimension space. 
i-DarkVec (Word2Vec) produces sender (word) embeddings such that senders (words) co-occurring in sequences (sentences) appear close in the latent space~\cite{gioacchini2023idarkvec}.

Formally, given a \textit{vocabulary} of senders $V=\{v_1,v_2,\ldots \}$ and the sequences of packets they send 
we map each entity ${v_j\in V \to u_j\in \mathbb{R}^E}$ where $u_j$ is the embedding of $v_j$ in the $E$ dimensional space. The function $e:V\to \mathbb{R}^E$ is the embedding function (\ie Word2Vec) that we train by giving in input the sequences of packets.

We consider each day as a new batch of data and we process it to extract the embedding of each day.

\subsection{Downtream Task: Unsupervised Clustering}
\label{ss:stage2}


We apply unsupervised downstream task on the produced embeddings using HDBSCAN~\cite{hdbscan}, a non-overlapping density-based clustering method. We employ the \textit{cosine distance} as the metric to measure the distance between sender embeddings.
Different from DBSCAN~\cite{ester1996density}, HDBSCAN simplifies the parameter selection. It only requires the \textit{minClusterSize}, an intuitive and easy-to-choose parameter that constrains the smallest size of clusters one wants. 

In short, HDBSCAN effectively identifies clusters of points that lie in a dense region of the space. It does not assume any shape for the cluster and accommodates clusters with diverse internal densities. It extracts well-shaped clusters, enhancing its robustness against noise. Points that fall in dense regions with less than \textit{minClusterSize} will be assigned the \textit{noise} cluster.

Formally, at each discrete time window $t$, we partition the senders $V^t$ into clusters $X_{i}^{t}$, $V^t=\bigcup_i X_{i}^{t}$. We call the partition $\zeta_t:=\{X_{i}^{t}\}_i$. The cluster $X_{-1}^{t}$ with the index $i=-1$ is the noise cluster.

We assess the cluster quality by measuring the \emph{Silhouette}, $Sh\in[-1,1]$, a measure of the cluster separation. Values close to 1 indicate perfect separation of clusters. An average silhouette of over 0.7 is considered to be strong, a value over 0.5 reasonable, and over 0.25 weak. Negative values are an indication of bad clustering.

Given the embedding obtained for each batch of data $t$, we cluster active senders to obtain $\zeta_t$.

\begin{figure*}
    \centering
        \begin{subfigure}[b]{.32\linewidth}
         \centering
         \includegraphics[width=\linewidth]{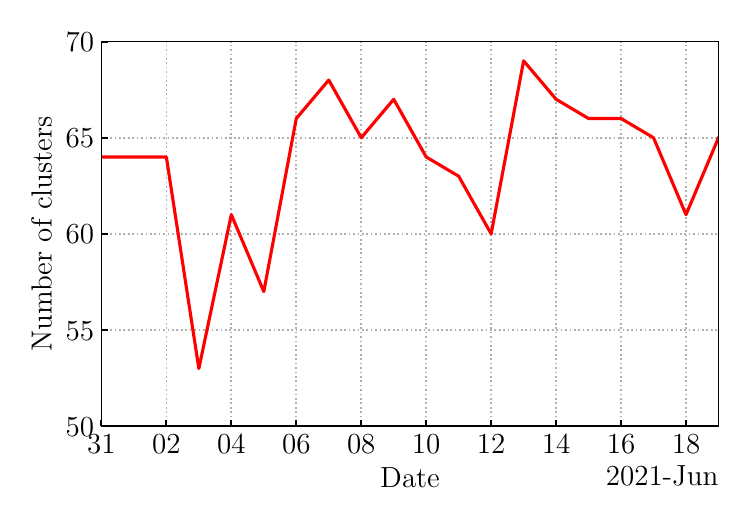}
         \caption{Number of detected clusters per day.}
         \label{fig:num_cluster}
    \end{subfigure}
    \begin{subfigure}[b]{.32\linewidth}
        \centering
        \includegraphics[width=\linewidth]{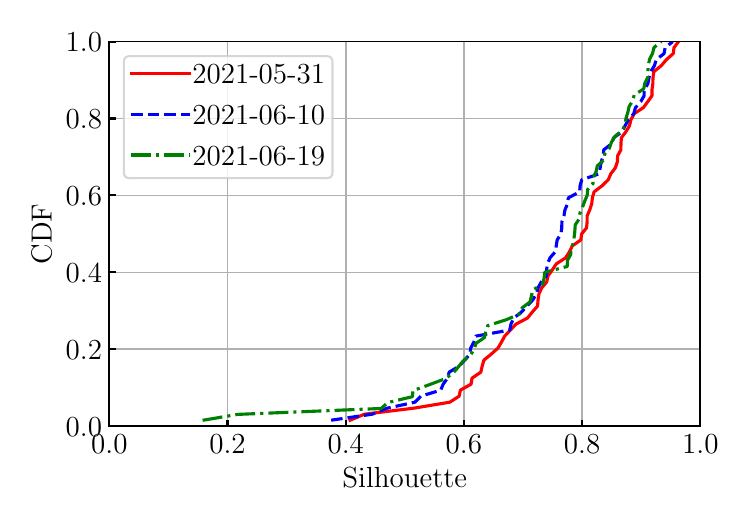}
        \caption{ECDF of silhouette values.}
        \label{fig:ecdf_sh}
    \end{subfigure}
    \begin{subfigure}[b]{.32\linewidth}
         \centering
         \includegraphics[width=\linewidth]{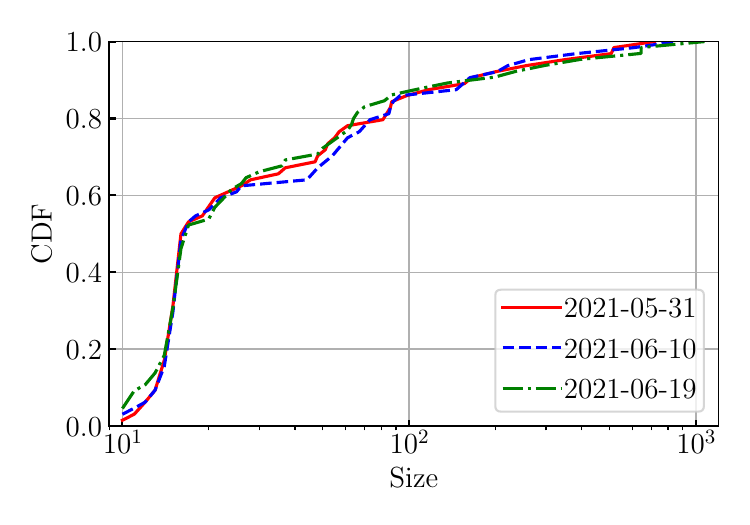}
         \caption{ECDF of cluster sizes.}
         \label{fig:ecdf_size}
    \end{subfigure}
    \caption{Statistics of clustering results.}
    \label{fig:clustering_results}
\end{figure*}

\subsection{Dynamic Clusters Analysis}
\label{ss:stage3}

At the end of each time batch $t$ we obtain an independent clustering $\zeta_t$. Hence, we need a way to track changes and detect novelties among the clusters, i.e., to solve the DCA problem.
MONIC~\cite{spiliopoulou2006monic} is a framework that implements a DCA to model and track cluster transitions. 
It takes as an input the clustering outcomes, 
e.g., $\zeta_t$ and $\zeta_{t+1}$ and defines a set of transitions that identify possible changes involving multiple clusters. 
Consequently, it offers valuable insights into the overall cluster dynamics.
To align the MONIC method with the IP embedding clusters scenario and to enhance its suitability, we introduce some modifications. This adaptation accounts for the presence of samples being clustered as noise and the significant variation in samples for clustering (i.e., the active senders) across different batches which are peculiar to our scenario.

Our DCA problem deals with the changes in members of clusters rather than their spatial properties.
Thus, the fundamental building block is the \emph{set-overlap} operator defined as
$OL(X, Y):=\frac{|X \cap Y|}{|X|}$, where $X$ and $Y$ are two sets. Notice that the overlap operator is not symmetrical.

Given two clustering outcomes $\zeta_t$ and $\zeta_{t+1}$, we identify various types of transitions according to the overlap among clusters. 
Let $X_i$ be a cluster in $\zeta_t$ and $Y_j$ be a cluster in $\zeta_{t+1}$.\footnote{For simplicity, we omit the index $t$ from the notation of the clusters.}
Different from MONIC, we include the noise cluster in the partitions, denoted as $X_{-1}$ and $Y_{-1}$.

Some senders active in $t$ may not be active in the following time slot, while some senders would still active in $t+1$. The fraction of still present senders is key to re-identify a cluster. 
Let us define the operator $A(X)$ as the number of \textit{active} elements of $X \in \zeta_t$ that are found in $t+1$: 
$$A(X) := {\sum_{Y\in\zeta_{t+1}} OL(X,Y)}.$$

Let $\tau_0 \in [0.5,1]$ and $\tau_1 \in (0,0.5)$ be two thresholds.

If $OL(X,Y)/A(X) \geq \tau_0$, we say that $X$ \textbf{strongly matches} $Y$.
If $ OL(X,Y)/A(X) \geq \tau_1$, we say that $X$ \textbf{loosely matches} $Y$.\footnote{This is a simplification w.r.t. MONIC original paper~\cite{spiliopoulou2006monic} in which there is an ageing function.}

First, we verify if a cluster is not anymore active in $t+1$:
\begin{itemize}
    \item \textbf{Inactive}: $A(X_i)
    < \tau_1$; i.e., the fraction of active senders in $t+1$ is smaller than $\tau_1$.
\end{itemize}
Otherwise, we say the cluster $X_i$ is \textbf{Active}.

For active clusters, we define the following transformations:
\begin{itemize}

    \item \textbf{Disappeared}: 
    $\forall Y_j\in\zeta_{t+1}: \frac{OL(X_i,Y_j)}{A(X_i)} < \tau_1$ or $\frac{OL(X_i,Y_{-1})}{A(X_i)} \geq \tau_0$; i.e., there is no loosely match with any other clusters; or there is a strong match with the noise cluster.

    \item \textbf{Split}:
     $\forall Y_j\in\zeta_{t+1}: \frac{OL(X_i,Y_j)}{A(X_i)} < \tau_0$ and $\exists Y_j\in\zeta_{t+1},j\neq -1:\frac{OL(X_i,Y_j)}{A(X_i)} \geq \tau_1$; i.e., there is not a strong match and there exists at least a loosely match in $t+1$.

    \item \textbf{Survived}: $\exists Y_j\in\zeta_{t+1},j\neq -1: \frac{OL(X_i,Y_j)}{A(X_i)} \geq \tau_0$
     and $\nexists k\neq i: \frac{OL(X_k,Y_j)}{A(X_k)} \geq \tau_0$; i.e., $X_i$ is the only cluster that strongly matches another cluster $Y_j\in\zeta_{t+1}$.
     Similarly, $X_i$ survives if $\exists k_2\neq i: \frac{OL(X_{k_2},Y_j)}{A(X_{k_2})} \geq \tau_0$ and $OL(Y_j,X_i) \geq \tau_0$; i.e., there are other clusters that strongly matches $Y_j$, but $X_i$ has a backward overlap larger than $\tau_0$ ($X_i$ is the main component of $Y_j$).

    \item \textbf{Absorbed}: $Y_j\in\zeta_{t+1},j\neq -1:\frac{OL(X_i,Y_j)}{A(X_i)} \geq \tau_0$ and
    $\exists k\neq i: \frac{OL(X_k,Y_j)}{A(X_k)} \geq \tau_0$; i.e., A cluster $X_i$ strongly matches the cluster $Y_j$, but there is also another cluster $X_k \in \zeta_{t}$  that strongly matches $Y_j$. in this case, we require that the backward overlap of $X_i$ w.r.t. $Y_j$ is limited, i.e. $OL(Y_j,X_i) < \tau_0$.

\end{itemize}

At last, we consider the case new clusters emerge. In details, we consider $Y_j$ in $\zeta_{t+1}$ as:
\begin{itemize}
    \item \textbf{Emerged}: $\forall X_i\in\zeta_{t},i\neq -1: \; \frac{OL(X_i,Y_j)}{A(X_i)} < \tau_0$; i.e.,  there is no cluster in $t$ that strongly matches $Y_j$.
    Or $\forall X_i\in\zeta_{t}: OL(Y_j, X_i) < \tau_0$; i.e., there are two or more clusters in $t$ that strongly matches $Y_j$, but none of them is a major component of $Y_j$.
    

    
\end{itemize}

The transitions Inactive and Emerged are not defined in the original formulation of MONIC. In addition, we added the backward overlap to make sure that if most of the points of $Y_j$ are present in $X_i$ then $X_i$ survived (otherwise it would have been absorbed).

In our dynamic cluster analysis, given a newly emerged cluster $Y$ at time $t+1$, we perform a backward matching also with the past batches, $b=[t-1, t-2, \ldots, 0]$.
For every past cluster $X \in \cup_b(\zeta_{b})$, we check whether there exists some old $X$ which overlaps $Y$, i.e.  if $OL(X,Y) \geq \tau_0$.
If there is not such a matching, we consider $Y$ as a \textbf{Novelty}, i.e. we observe $Y$ for the first time in $t+1$.

\begin{table*}
\centering
\footnotesize
\caption{Cluster evolution on the 20 days of test according to MONIC framework.}
\begin{tabular}{l||cccc||cccc}
\toprule
 & \multicolumn{4}{c||}{\textbf{Labelled}} & \multicolumn{3}{c}{\textbf{Unknown}} \\
 & \textbf{Clusters$^{\dagger}$} & \textbf{Avg. IPs} & \textbf{Avg. Sh} & \textbf{Avg. purity} & \textbf{Clusters} & \textbf{Avg. IPs} & \textbf{Avg. Sh} \\
\midrule
Absorbed & 3.5\% & 110  & 0.56 & 0.89 & 2.6\% & 69 & 0.63 \\
Split & 2.1\% & 186 & 0.47 & 0.75 & 2.8\%& 72 & 0.60\\
Disappeared & 2.0\% & 19 & 0.70 & 0.93& 3\% & 16 & 0.69\\
Survived & 85.2\% & 41 & 0.79 & 0.95 & 71.8\% & 119 & 0.73\\
Inactive & 7.2\% & 38 & 0.86 & 0.96& 19.7\% & 174 & 0.72 \\
\midrule
Total & 785 & 46 & 0.78 & 0.95 &490 & 125 & 0.72\\
\midrule\midrule
Emerged$^{\dagger\dagger}$ & 115 & 74 &  0.70& 0.90 &127 & 124 &  0.71\\
\bottomrule
\multicolumn{8}{l}{\begin{tabular}[c]{@{}l@{}}$^\dagger$ Percentages exclude the clusters on the last day, since we do not know their future transitions.\end{tabular}}\\
\multicolumn{8}{l}{\begin{tabular}[c]{@{}l@{}}$^{\dagger\dagger}$  We detail characteristics of Emerged clusters (already included in the Total row).\end{tabular}
} 
\end{tabular}
\label{tab:transitions}
\end{table*}

\section{Ground Truth for Testing}
\label{s:dataset}



To verify the performance of both the clustering and the dynamic analysis we leverage a partial ground truth (GT) representing 13 classes of senders whose coordination is known a priori. These are mostly set of IP addresses belonging to well-known projects or botnets.
In detail, we label Mirai-like senders by checking the Mirai packet signature~\cite{mirai_fingerprint}, and we label acknowledged internet scanners publicly advertised in online repositories\footnote{\url{https://gitlab.com/mcollins_at_isi/acknowledged_scanners}}. 
We mark all the remaining senders as \emph{Unknown}, ending up with a highly unbalanced GT -- \eg thousands of senders exhibit the Mirai-like label, while only hundreds or dozens belong to classes of acknowledged scanners.

To assign a cluster label, we employ a majority voting rule, \ie we assign a cluster the most frequently occurring label within samples in that cluster. We also consider the \textit{purity} of a cluster\cite{schutze2008introduction}, \ie the proportion of samples within a cluster bearing the predominant label.

In general, we expect that senders belonging to the same GT class shall consistently belong to the same cluster which survives over time. We eventually expect subgroups of senders in a given class to perform different scan activities\footnote{We already observed subsets of senders of a known class that perform different scans~\cite{gioacchini2023idarkvec}.}. We instead expect new clusters to Emerge, Disappear or be Absorbed for senders in the unknown class.

\section{Experimental Results}
\label{s:results}
The size of the embeddings $E$ is chosen as 200, as in \cite{gioacchini2023idarkvec}.
We set HDBSCAN with \textit{minClusterSize} = 10 and $\tau_0=0.65$ and $\tau_1=0.3$ for the DCA algorithm. While there is room for future tuning of these parameters, we empirically found these choices have little impact on the overall results.

\subsection{Generic Clustering Results}

As previously said, we consider batch size equal to one day and run HDBSCAN for the 20 testing days in our dataset. For each day, we apply clustering only for active senders.

Figure~\ref{fig:num_cluster} shows the number of clusters found in each day. We observe a quite variable setup, with from 53 to 69 clusters found each day. Considering the quality of clustering, in Figure~\ref{fig:ecdf_sh} we show the empirical cumulative distribution function (ECDF) of the silhouette for the clusters identified in the initial, mid, and final days of the dataset. We observe some variations of cluster quality across different dates, but typically more than 80\% of clusters have $sh\geq0.6$, i.e., most of the clusters are well-separated and compact. To complete the clustering characterization,  Figure~\ref{fig:ecdf_size} shows the ECDF of cluster sizes. Most of them have less than 100 senders but some tops to 1,000 samples (notice the log x-scale).

In short, the clustering step produces few and well-shaped clusters, reducing by two order of magnitude the entities to analyse. The question remains which of the cluster present at batch $t+1$ can be linked to clusters seen in the past batches. We answer to this question next.

\begin{figure}
    \centering
    \begin{subfigure}[b]{.8\linewidth}
        \centering
        \includegraphics[width=\linewidth]{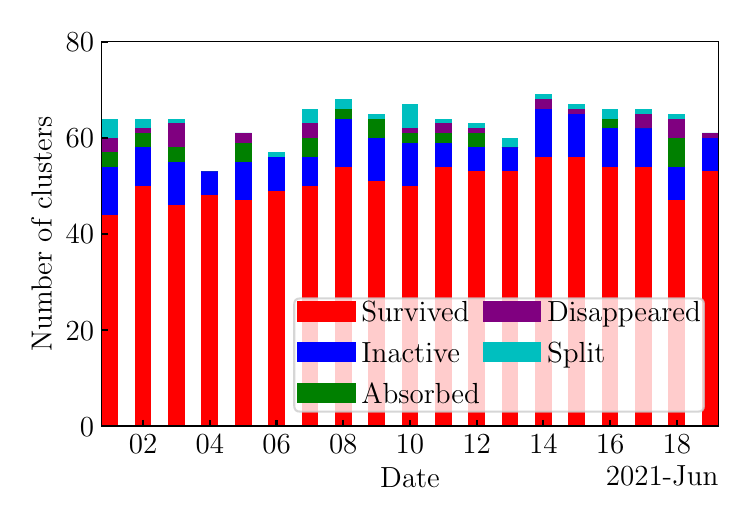}
        \caption{Cluster transitions per day.}
        \label{fig:trans_cluster}
    \end{subfigure}
    \begin{subfigure}[b]{.8\linewidth}
         \centering
         \includegraphics[width=\linewidth]{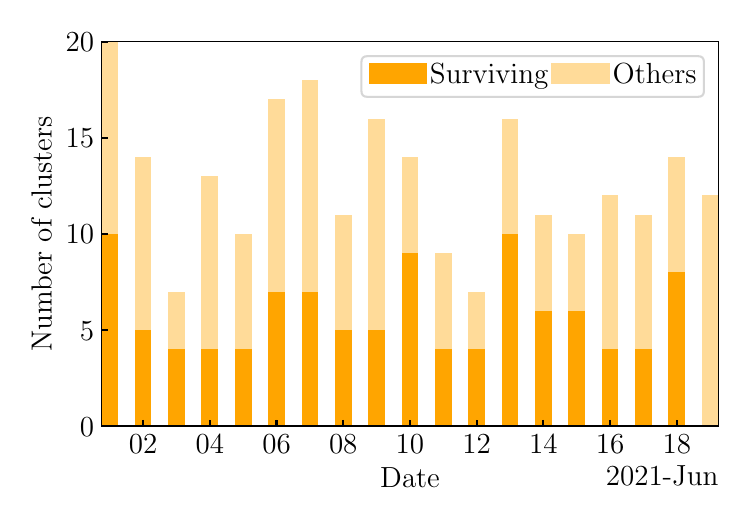}
         \caption{Emerged clusters per day.}
         \label{fig:emerged_clusters}
    \end{subfigure}
    \caption{Dynamic cluster analysis.}
    \label{fig:transitions}
\end{figure}

\begin{figure}[!ht]
\centering
    \includegraphics[width=.8\linewidth]{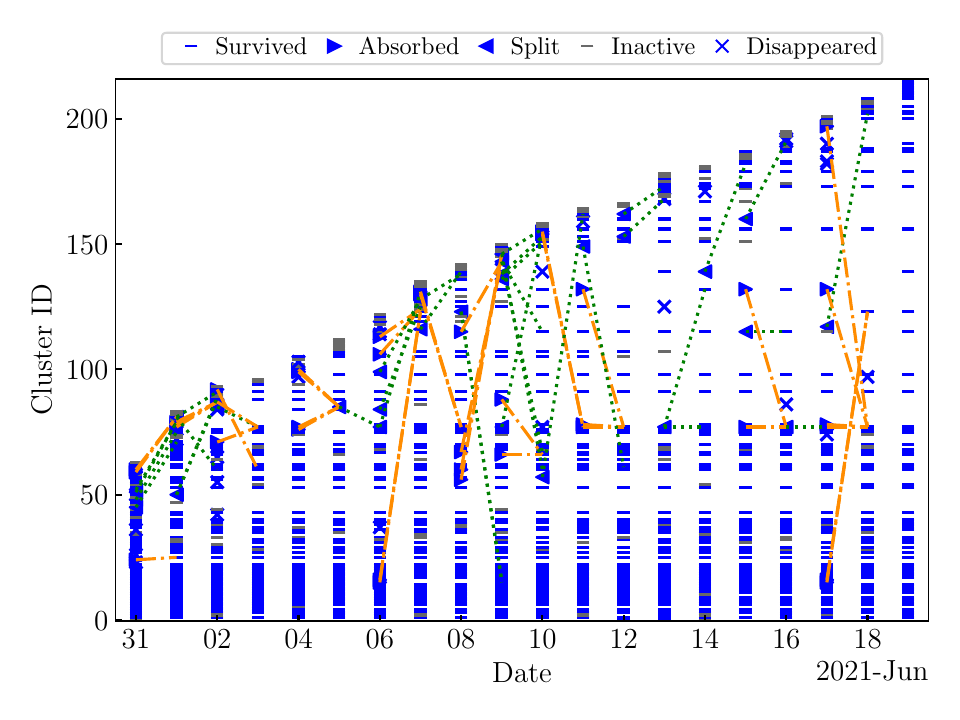}
    \caption{Evolution of all activities. Lines link clusters that split (dotted green line) or get absorbed (orange dotted line).}
    \label{fig:generic_picture}
\end{figure}

\begin{figure*}
    \centering

    \begin{subfigure}[b]{.45\linewidth}
         \centering
         \includegraphics[width=\linewidth]{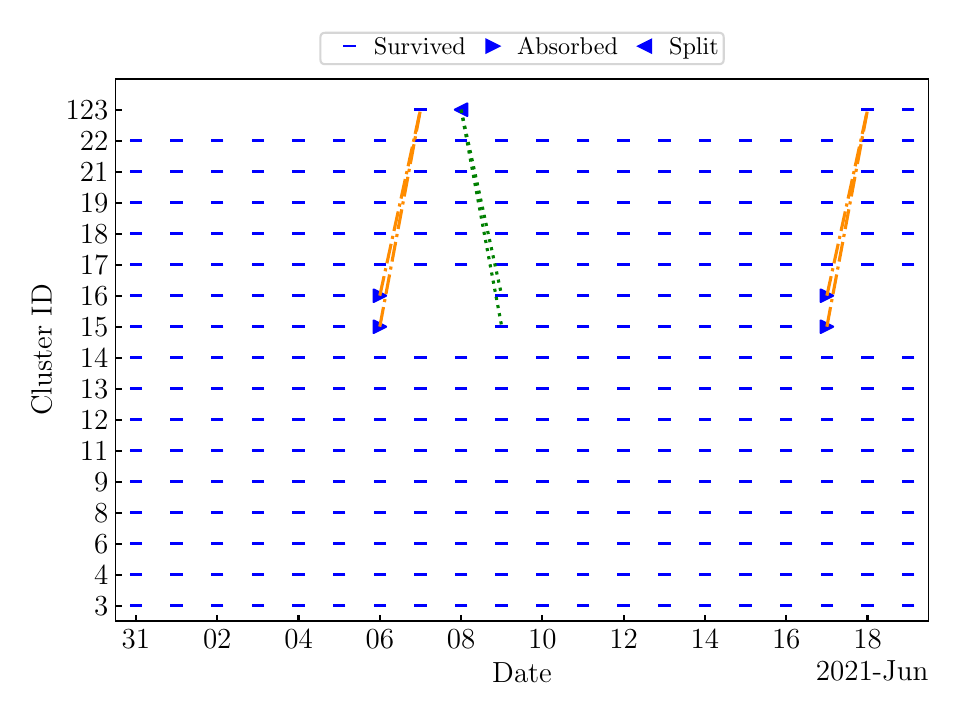}
         \caption{Shadowserver clusters.}
         \label{fig:shadowserver}
    \end{subfigure}
    \begin{subfigure}[b]{.45\linewidth}
         \centering
         \includegraphics[width=\linewidth]{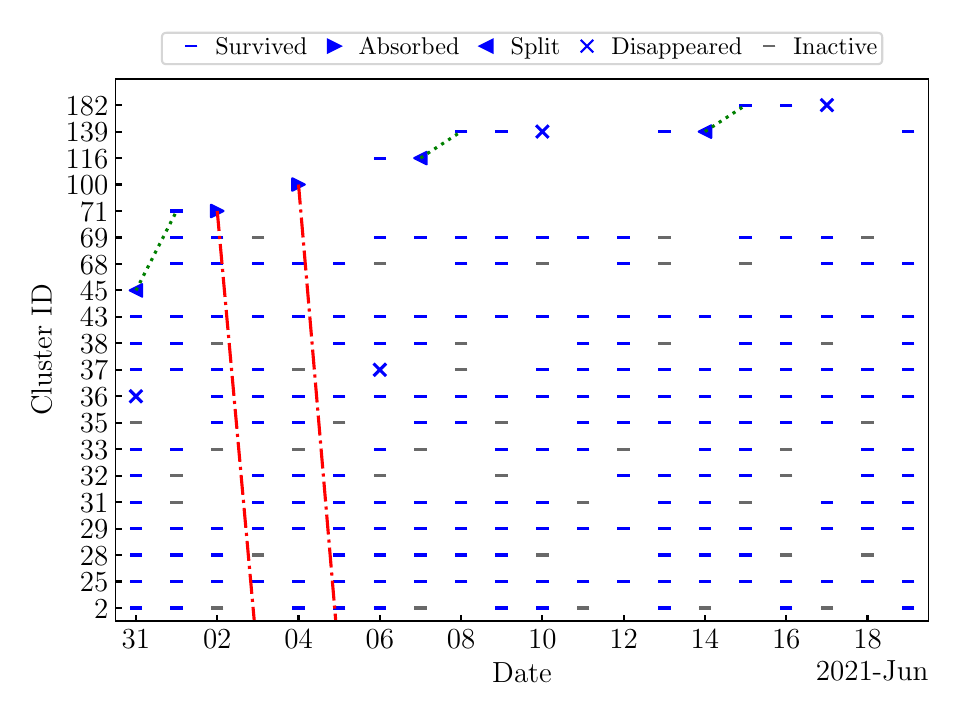}
         \caption{Censys clusters.}
         \label{fig:censys}
    \end{subfigure}
    \caption{Examples of evolution of some clusters of known classes with DCA framework.}
    \label{fig:examples_2_classes}
\end{figure*}

\subsection{Overall Results of DCA}
In Figure \ref{fig:transitions}, we evaluate cluster evolution with the outcomes of the DCA. 
Every day, after obtaining new clustering, we backtrack changes in activities from the previous days. Figure~\ref{fig:trans_cluster} shows the breakdown of clusters among Survived, Inactive, Absorbed, Disappeared and Split. We observe that the majority of clusters (40 to 50) survive, indicating most of the previously identified coordinated activities persist. There is no need to analyse these new clusters if one did it in the past.

A small portion of clusters becomes inactive (blue), i.e., their senders are not present in $t+1$ anymore. Very few clusters occurred in some other transition, most of which correspond to senders being Absorbed by larger clusters (green) or getting Split (cyan), i.e., their behaviour is not part of a strong coordinated pattern. By manually inspecting some of these cases, we observe mostly noisy senders which send few packets but do not follow a particular pattern.

Each day, the DCA algorithm identifies also some Emerged clusters (7-20 each day). Figure~\ref{fig:emerged_clusters} details this reporting those that will Survive in the next days and those that will never be re-identified in the future. About half of them would survive for at least one more day. The analysts should give priority to checking these potentially new (and persistent) clusters that could represent new anomalies and threats. 

Using the DCA we identify 216 different 
coordinated activities, we can label about 37.5\% of them. Figure~\ref{fig:generic_picture} provides an overview of the overall pattern of these.
On the x-axis, we have the time. The y-axis reports the cluster ID (in increasing order). When a new cluster Emerges for the first time, we assign it a new ID. If a cluster Survives, we report it by assigning the previous ID and reporting it with a dash. When a cluster disappears, we mark the last time we saw it by a corresponding ending event (Absorbed or Disappeared). In the case of Split, we link the new clusters to the originating one (orange lines). Same for Absorbed (green lines).

Some consideration holds: 1) some clusters are identified on the first day and survive throughout the 20 days (bottom part of the plot); 2) some emerge later and then survive for multiple days (linear growth followed by horizontal dashes); 3) some become inactive for several days, then re-appear (intermittent dash patterns). Only a few clusters Split or are Absorbed.

For completeness, we detail the transitions with labelled and unknown clusters in Table~\ref{tab:transitions}. As expected, clusters with labels are more stable because they involve senders controlled by research projects or security crawlers that exhibit regular and constant activities. The DCA can easily match most of them as Surviving (85.2\%). Notice 115 clusters Emerge\footnote{We do not consider the clusters at the first day as Emerged.}, a small fraction compared to the 785 that we observe in the whole 20 days.
Notice that these labelled clusters exhibit high purity and silhouette, which confirms accurate clustering. A small portion of clusters have low purity and tend to Split more easily. For instance, some noise points are initially clustered together labelled points. In the next batch, the clustering step may remove those noise points (which may generate a separate cluster, resulting in a Split). 

Focus now on unknown clusters -- right part of the Table~\ref{tab:transitions}. They manifest greater dynamism with nearly 20\% of them becoming inactive (w.r.t. 7.2\% for labelled clusters). Only 72\% of clusters survive for more than one day (w.r.t. 85\%). We also witness more Emerging clusters (127 Emerged over a total of 490 clusters identified over 20 days). Interestingly, unknown clusters are generally larger than labelled ones. This can be linked to possible botnet activities which can involve a larger number of senders than research projects (which rely on limited and fixed senders acting as scanners). In fact, the diversity of IP addresses and subnets found in these clusters confirms this assumption.

\subsection{Example of Labelled Cluster Evolution}
To give the intuition of the goodness of the DCA process, we focus on the clusters of senders belonging to two well-known acknowledged security engines: Shadowserver~\cite{shadowserver} and Censys~\cite{censys}. We know these projects keep periodically scanning the entire IPv4 address space looking for vulnerable services. They generate a continuous and regular stream of packets and we expect them to be consistently present and well-identified by i-Darkvec and the clustering step, while we expect the DCA to keep strongly matching current clusters with past ones (Survived).

Focus on Shadowserver first. In Figure~\ref{fig:shadowserver} we show the evolution of labelled clusters with the majority of senders being known Shadowserver scanners. 
We identify 16 clusters on the first day (we report them using their unique cluster ID as it appears in Figure~\ref{fig:generic_picture}). This reflects the fact that different groups of Shadowserver's senders perform different scan activities.
In the following days, the DCA analysis shows a very stable picture which reflects the periodic scan Shadowserver's senders do -- \ie all of the clusters Survive for 20 days. There are a few exceptions: clusters 9 and 10 which are absorbed together into cluster 16 for some days. By manually verifying these cases, we notice in fact that senders in these two clusters generate less traffic on those days, and our pipeline reflects this. 
\begin{figure*}
    \centering
    \begin{subfigure}[b]{.45\linewidth}
         \centering
         \includegraphics[width=\linewidth]{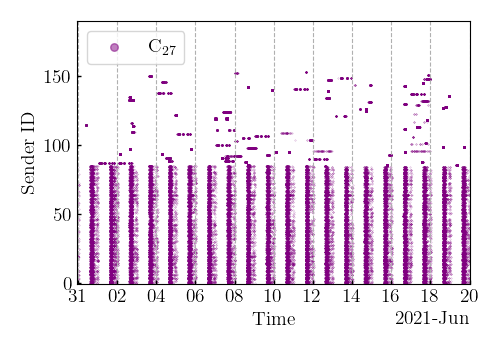}
         \caption{Online services scanning cluster.}
         \label{fig:services}
    \end{subfigure}
    \begin{subfigure}[b]{.45\linewidth}
         \centering
         \includegraphics[width=\linewidth]{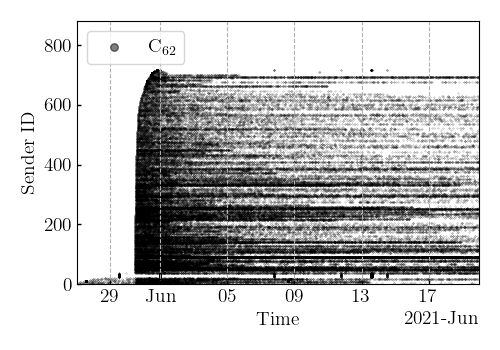}
         \caption{GRE backscattering cluster.}
         \label{fig:c62}
    \end{subfigure}
    \begin{subfigure}[b]{.45\linewidth}
         \centering
         \includegraphics[width=\linewidth]{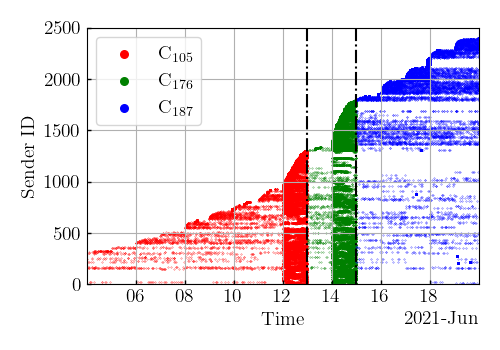}
         \caption{9530/TCP-scan clusters evolution.}
         \label{fig:9530tcp}
    \end{subfigure}
    \caption{Patterns of activity for senders of some example clusters.}
    \label{fig:patterns}
\end{figure*}

Focus now on Censys in Figure~\ref{fig:censys}. We identify 20 clusters, all labelled as Censys. Here, some clusters become Inactive (missing dash) and reappear from time to time. Some are Absorbed into clusters with unlabelled IPs (red dashed lines in the plot), Split or even Disappear into noise. By manually checking some of these events, we confirm that Censys scanners are split into groups performing different scan activities. Their activity is less constant in time. In fact, some senders sometimes disappear for days or portions of a day. This alters the sequence of packets i-Darkvec sees, modifying the host embedding and making these more similar to (and thus clustered with) other intermittent or noisy senders. The DCA pinpoints these changes. When those senders become active again, i-Darkvec and the clustering assign them to their original cluster and the DCA matches it with its last occurrence.

These two examples show the robustness of our DCA methodology in successfully clustering and matching senders performing similar activities.



\section{Novelty Examples}

We manually analyse some of the clusters (mostly unknown) and look for insights about their activities. We leverage external information offered by DNS and whois and summarise some interesting findings.

\subsection{Detailed Examples}

\textbf{\textbf{SIP and RDP scanners in Google Cloud}} Figure \ref{fig:services} shows the patterns of senders in Cluster 27 that DCA identifies in the first day. The $x$-axis represents the time, and the $y$-axis the senders, ordered by time of appearance; a point marks the time in which the given sender sends a packet. 
The bottom 90 senders are from three /24 subnets located in different Google Cloud IP address ranges. Their activities show evident temporal patterns (see the bottom of the figure), which persist for 20 days. These senders target port 5060/TCP (SIP initialization protocol) and port 3389/TCP (used for Windows Remote Desktop - RDP). {Some of them are reported in some database 
for unauthorized connection attempts}. This evidence strongly suggests some malicious scan activity from actors leveraging the Google Cloud. Senders 90-150 in the top part of the Figure are less regular but perform scans to the same ports with a similar pattern. Thus the clustering algorithm correctly places them in the same cluster (C$_{27}$). Their IP are like bots scattered in random networks and also reported for abuse in some database. These senders could be controlled by other entities, or be part of a distributed botnet the malicious actor controls as well.

\textbf{\textbf{GRE DDoS attack}} Figure \ref{fig:c62} shows the activity pattern exhibited by Cluster 62 (C$_{62}$) senders over time. The $y$-axis shows the senders of this cluster in order of appearance. 
Here we are also reporting ($x$-axis) what happens during some prior days of bootstrapping.  This cluster suddenly emerges and survives for 20 days. In the first days, we observe more than 700 senders. The number keeps shrinking, with about 100 senders still present on the last day. Around 30\% of the senders belong to the domain name \url{rr.com}, a webmail provider. Most of the traffic is Generic Routing Encapsulation (GRE) protocol. Senders hit our network telescope on 2021-05-30, right before our test period and become very active in the following few days. Manually checking the packet trace, we conclude these are back-scattering packets sent by victims of a Distributed Denial of Service (DDoS) attack where \url{rr.com} senders are the victims. The attacker sends GRE packets to \url{rr.com} servers with randomly spoofed IP sender addresses. By chance, these happen to be one of our /24 darknet addresses. The victim then replies with a GRE error message, sending packets to our darknet IP addresses. The decrease in senders is likely due to some mitigation action the victim put in place.

\textbf{\textbf{9530/TCP scan clusters}} Figure \ref{fig:9530tcp} shows the evolution of Clusters 105, 176 and 187.  These senders mostly target port 9530/TCP, which is reported as a vulnerability (backdoor) of some surveillance devices~\cite{habr2024}.  We identify C$_{105}$ firstly on 2021-06-03 (first day shown on the $x$-axis) with only 15 senders (11 of them are Mirai-like senders). This cluster expands significantly up to about 800 senders on 06-12 when suddenly we observe a lot of packets (see the dense red pattern). Most of those senders go suddenly silent the day after, and only a small portion remains still active. The DCA fails to match the new cluster C$_{176}$ with C$_{105}$ (because $OL(C_{105},C_{176})/ A(C_{105}) < \tau_0$) as most senders of C$_{105}$ are inactive now). We thus identify C$_{176}$ as newly emerged and classify it as a novelty. Most of the inactive senders suddenly become active again in 06-14 and are now matched with C$_{176}$ (see the dense green pattern). On 06-15, they suddenly become inactive, and the DCA identifies the few remaining active as a new cluster -- C$_{187}$, whose members grow in the next few days.

This example shows our approach successfully identifies the dynamic. Yet, the extreme on/off/on pattern challenges our ability to match newly emerged clusters with previous ones. This shows the complexity of possible scenarios, and opens the opportunity to improve the DCA.

\subsection{Other Examples}

\textbf{\textbf{Mongolian /24 scanner (C$_0$)}} The cluster appears for the first time on 2021-05-31 then became inactive for several days until the 2021-06-13 when it survives for three days. In total, we observe 256 senders in this cluster, 253 of which belong to the same subnet 180.149.126.0/24
located in Mongolia. These senders perform TCP scans to some very specific ports for the whole 48 hours of the weekend in local time. 
Some of these addresses have been reported in the past as a possible network scanner\footnote{\url{https://www.abuseipdb.com/check/180.149.126.40}}.

\textbf{\textbf{Targeted port scan (C$_7$)}} The cluster survives throughout the 20 days. The cluster is composed of 21 senders distributed in random networks (only four of them are from the same /24 subnet). All senders were performing scans at a high frequency for 20 days. However, on the evening of 06-07 all scan activities paused for about 12 hours, which reflects an evident coordination. Senders constantly target the same 35 ports, all well-known for vulnerable services, \eg port 1900 used by SSDP\cite{ssdp}. This clearly hints at a malicious coordinated activity. Even if only 21 senders are involved, our methodology could pinpoint and highlight it. 

\textbf{\textbf{New Shadowserver senders (C$_{20}$)}} The cluster survives for the whole 20 days, being remarkably stable. It includes about 200 senders from the same /24 network {64.62.197.0/24}. A reverse DNS lookup reports these belong to Shadowserver, and their temporal patterns are consistent with the other known Shadowserver scanners. Hence, we confirm a new group of Shadowserver senders was added to their infrastructure, but not included in the publicly available GT.

\textbf{\textbf{Botnet targeting port 23 (C$_{61}$)}} The cluster is quite large and dynamic. Its size varies dramatically day by day, from around 500 to more than 1,100. Each day, about half of the senders in that cluster became inactive and hundreds of new senders joined. Nevertheless, the cluster is always identified as surviving. In total, we observe nearly 7 thousand senders in the cluster, with more than 3,500 IPs just appearing once. They target mostly the 23/TCP port. This suggests these senders are part of a large botnet. Interestingly, on the first day, 90\% of senders are labelled Mirai-like. The cluster purity decreases from 0.9 to about 0.5 so that since day 13 the majority of senders are unknown. After inspecting the traffic of these new senders, we find that most of them exhibit the Mirai fingerprint. These are new senders that were not present in the dataset we used to build the GT. This testifies the Mirai-like botnet keeps growing, and that our approach can simplify the tracking of new bots appearing.
















\section{Conclusions}
\label{s:conclusion}
In this paper, we presented a preliminary exploration to track the evolution of coordinated activities and detect novelty from traffic observed via network telescopes. We proposed a dynamic clusters analysis approach for activities in network telescopes inspired by the MONIC framework.

Experimental results show that our methodology can: (i) reduce human effort to analyse newly observed groups of coordinates senders; (ii) provide deeper insights into activities, highlight changes and incidence by tracking the 
clusters evolution.

Future development includes improving the robustness of the approach by defining improved strategies to match clusters at distant times and thoroughly exploring the sensitivity to parameters. A promising extension of the proposed methodology is to leverage other features in addition to those extracted by i-DarkVec, introducing criteria on priority to check new clusters and automatically highlight changes within a cluster. Furthermore, we are looking forward to deploying the system in real-time and performing continuous analysis.

\section*{Ethical Considerations}
Our work focuses solely on technical advancements in utilising dynamic cluster analysis to enhance cybersecurity and deepen the knowledge of traffic observed in network telescopes. We only employ passive measurement techniques, this means that we do not engage with or influence any entities we measure.
No further ethical considerations apply to our research, as our primary objective is to improve the effectiveness of network security measures through innovative AI solutions.

\section*{Acknowledgements}
This work has been partly funded by the project SERICS (SEcurity and RIghts In the CyberSpace - PE00000014) under the MUR National Recovery and Resilience Plan funded by the European Union, and the xInternet (eXplainable Internet - 20225CETN9) projects - funded by European Union - Next Generation EU within the PRIN 2022 program (D.D. 104 - 02/02/2022 Ministero dell'Università e della Ricerca). This manuscript reflects only the authors' views and opinions and the Ministry cannot be considered responsible for them.

\bibliographystyle{plain}
\bibliography{bibliography}


\end{document}